\documentclass[10pt]{article}
\usepackage{latexsym}
\usepackage{amsthm}
\usepackage{mathrsfs}
\usepackage{amsfonts}
\usepackage{enumerate}
\usepackage{amsmath}
\usepackage{amssymb}
\begin{document}
\begin{center}
\null\vspace{2cm}
{\large {\bf Hawking radiation of Reissner-Nordstr\"{o}m-de Sitter black hole by Hamilton-Jacobi method}}\\
\vspace{2cm}
M. Ilias Hossain\footnote{E-mail: $ilias_-math@yahoo.com$}\\
{\it Department of Mathematics, Rajshahi University,  Rajshahi - 6205, Bangladesh}\\
M. Atiqur Rahman\footnote{E-mail: $atirubd@yahoo.com$}\\
{\it Department of Applied Mathematics, Rajshahi University, Rajshahi - 6205, Bangladesh}\\

\end{center}
\vspace{3cm}
\centerline{\bf Abstract}
\baselineskip=18pt
\bigskip

 In Refs. (M. Atiqur Rahman, M. Ilias Hossain (2012) Phys. Lett. B {\bf 712} 1), we have developed Hamilton-Jacobi method for dynamical spacetime and discussed Hawking radiation of Schwarzschild-de Sitter black hole by massive particle tunneling method. In this letter, we have investigated the hawking purely thermal and nonthermal radiations of Reissner-Nordstr\"{o}m-de Sitter (RNdS) black hole. We have  considered energy and angular momentum as conserved and shown that the tunneling rate is related to the change of Bekenstein-Hawking entropy and the derived emission spectrum deviates from the pure thermal spectrum. The results we have obtained for RNdS  black hole is also in accordance with Parikh and Wilczek\rq s opinion and recovered the new result for Hawking radiation of RNdS black hole.
\vspace{0.5cm}\\
{\bf Keywords: Massive Particle Tunneling, RNdS black hole.}\\
\vfill

\newpage

\section{Introduction}\label{sec1}
 A wonderful fact of black hole radiation  \cite{one,two} have discovered by Hawking in $1975$ and several works have been done to calculate this quantum effect \cite{three}. Nowadays, the radiation of black holes is called `Hawking radiation'. Furthermore Hawking proposed that the radiation of black holes can be shown as tunneling and the emission spectrum in light of quantum field theory in curved spacetime with the exception of following the tunneling picture. The tunneling phenomenon has been extensively studied \cite{four, five, six,seven,eight,nine,ten,eleven,twelve} and a lot of work has already been successfully applied on various black hole spacetimes  \cite{thirteen,fourteen,fifteen,sixteen,seventeen,eighteen,nineteen,twenty,twenty one,twenty two,twenty three,twenty four,twenty five,twenty six,twenty seven,twenty eight,twenty nine,thirty,thirty one,thirty two,thirty three, thirty four,thirty five,thirty six,thirty seven,thirty eight,thirty nine, fourty, fourty one, fourty two, fourty three} where a particle moves in dynamical geometry and all of these works are limited to massless particle and gives a correction to the emission rate arising from loss of mass of the black hole crresponding to the energy carried by radiated quantum. The method delineated Hawking radiation as tunneling process was first disclosed by Kraus and Wilczek \cite{fourty four, fourty five} and then reinterpreted  by Parikh and Wilczek \cite{fourty six}.  In this method the tunneling rate is related to the calculating of the imaginary part of the action for the process of s-wave emission across the horizon, which in turn is related to the Boltzmann factor for emission at the Hawking temperature. In general, based on semiclassical tunneling picture two universal methods are applied in references to derive the action. One method is called as the Null Geodesic method developed by Parikh and Wilczek \cite{fourty four,fourty five} and another method, proposed by Angheben et al. \cite{fourty seven} known as Hamilton-Jacobi methods and it is an extension of the complex path analysis proposed by Padmanabhan et al. \cite{fourty eight, fourty nine,fifty}.

In 2005, Zhang and Zhao have proposed the Hawking radiation from massive uncharged particle tunneling \cite{fifty one} and charged particle tunneling \cite{fifty two}. Following this work several researches have been carried out as charged particle tunneling \cite{fifty three,fifty four, fifty five,fifty six}. Kerner and Mann have developed quantum tunneling methods for calculating the thermal radiation spectrum of Taub-NUT black holes \cite{fifty seven} using both the null-geodesic and Hamilton-Jacobi methods by ignoring the self-gravitation interaction and energy conservation of emitted particle. However, according to the Parikh and Wilczek opinion, the radiation spectrum is not strictly thermal but satisfies the underlying unitary theory when self-gravitation interaction and energy conservation are considered. Considering Kerner and Mann\rq s process Chen, Zu and Yang reformed Hamilton-Jacobi  method for massive particle tunneling and investigate the Hawking radiation of the Taub-NUT black hole \cite{fifty eight}. Using this method Hawking radiation of Kerr-NUT  black hole \cite{fifty nine}, the charged black hole with a global monopole \cite{sixty} have been reviewed.

Recently, Atiqur and Ilias have reformed Hamilton-Jacobi method and investigate the Hawking radiation of the SdS black hole \cite{sixty one} where the position of the black hole horizon is taken in a series of black hole\rq s parameters so that the spacetime metric becomes dynamical and self-gravitation interaction  are taken into account. Here, we also assume that the changed of background geometry can be treated as the loss of radiated energy of the black hole. In this letter, the same method have been applied to investigate the Hawking radiation of Reissner-Nordstr\"{o}m-de Sitter (RNdS) black hole. In order to narrate Hawking-Radiation from the action of radiation particles the  method of Chen et al. \cite{fifty eight} is used. Our chief purpose concerned of this work is to calculate the imaginary part of action from Hamilton-Jacobi equation avoid by exploring the equation of motion of the radiation particle in Painlev\'e coordinate system and calculating the Hamilton equation. Though the equation of motion of massive particles are different from massless particle, We no need differentiate radiation particle. Above all as the self-gravitational interaction and the unfixed background spacetime are not assumed, the derived radiation spectrum deviates from the purely thermal one and the tunneling rate is related to the change of Bekenstein-Hawking entropy.

The cosmological constant with positive sign plays a prominent role in two reasons. First, the accelerating expansion of our universe indicates the cosmological constant might be a positive one \cite{sixty two,sixty three,sixty four}. Secondly, conjecture about de Sitter/CFT correspondence \cite{sixty five,sixty six} has been suggested that there is a dual relation between  quantum gravity on a dS space and Euclidean conformal field theory (CFT) on a boundary of dS space  \cite{sixty seven}. The outgoing particles tunnel from black hole horizon and incoming particles tunnel from cosmological horizon and formed Hawking radiation and the incoming particles can fall into the horizon along classically permitted trajectories for black hole horizon, but outgoing particles can fall classically out of the horizon for cosmological horizon.

This paper is arranged as follows. The later section describes the RNdS black hole spacetime with the position of event horizon. Near the event horizon the new line element of RNdS black hole is also derived here. The unfixed background spacetime and the self-gravitational interaction are taken into account, we review the Hawking radiation of RNdS black hole from massive particle tunneling method in section 3. In section 4, we have developed the Hawking purely thermal rate from non-thermal rate. Finally, in section 5, we present our remarks.

\section{Reissner-Nordstr\"{o}m-de Sitter black hole}\label{sec2}
The line element of Reissner-Nordstr\"{o}m-de Sitter black hole describing a charged, asymptotically de Sitter black hole with positive cosmological constant $\Lambda(=3/\ell^2)$ term is given by
 \begin{eqnarray}
ds^2 =-f(r)dt^2+\frac{1}{f(r)}dr^2+r^2(d\theta^2+\textrm{sin}^2\theta d\phi^2),\label{eq1}
\end{eqnarray}
where
 \begin{eqnarray}
f(r)=1-\frac{2m}{r}-\frac{r^2}{\ell^2}+\frac{q^2}{r^2}, \label{eq2}
\end{eqnarray}
 $m$ being the mass, $\ell$ is the cosmological radius, $q$ the total charge(electric plus magnetic) with respect to the static de Sitter space are defined such that
 $-\infty\leq t\leq \infty $, $r\geq 0$, $0\leq \theta \leq
\pi $, and $0\leq \phi \leq 2\pi$. At large $r$, the metric (\ref{eq1}) tends to the dS space limit. It is seen that the explicit dS case is
obtained by setting $m=0$ while the explicit Reissner-Nordstr\"{o}m case is obtained by taking the limit $\ell\rightarrow\infty$ and if we set $-\ell^2$ in the place of  $\ell^2$, the metric (\ref{eq1}) describes an interesting nonrotating AdS black hole called the Reissner-Nordstr\"{o}m Anti-de Sitter (RNAdS) black hole.

The black hole parameters M, q, and $\ell$ are related to the roots of
$r^4-\ell^2 r^2+2m\ell^2 r-\ell^2q^2=0$. Solving this equation by using Matlab or Mthematica we get the black hole (event) horizon $r_h$ and the cosmological horizon $r_c$ are respectively, at
\begin{eqnarray}
r_h&=&\frac{\ell}{\sqrt{3}}\textrm{sin}\bigg[\frac{1}{3}\textrm{sin}^{-1}\frac{3m\sqrt{3}}{\ell\sqrt{1+\frac{4q^2}{\ell^2}}}\bigg]\times
\bigg(1+\sqrt{1-\frac{q^2 \ell}{\sqrt{3} m}. \frac{2}{1+\delta}\textrm{cosec}\bigg[\frac{1}{3}
\textrm{sin}^{-1}\frac{3m\sqrt{3}}{\ell\sqrt{1+\frac{4q^2}{\ell^2}}}\bigg]}\bigg)\label{eq3},\\
r_c&=&\frac{\ell}{\sqrt{3}}\textrm{sin}\bigg[\frac{1}{3}\textrm{sin}^{-1}\frac{3m\sqrt{3}}{\ell\sqrt{1+\frac{4q^2}{\ell^2}}}\bigg]\times
\bigg(\sqrt{1+\frac{(1+\delta)\ell{3m}}{2\sqrt{3}}\textrm{cosec}^{3}\bigg[\frac{1}{3}
\textrm{sin}^{-1}\frac{3m\sqrt{3}}{\ell\sqrt{1+\frac{4q^2}{\ell^2}}}\bigg]}-1\bigg),\label{eq4}
\end{eqnarray}
where
\begin{equation}
\delta=\sqrt{1-\frac{4q^2}{3m^2} \textrm{sin}^2\bigg[\frac{1}{3}
\textrm{sin}^{-1}\frac{3\sqrt{3}m}{\ell\sqrt{1+\frac{4q^2}{\ell^2}}}\bigg]}\label{eq5}
\end{equation}
Expanding  $r_h$ in terms of $m$, $\ell$ and $q$ with $27\frac{m^2}{\ell^2}<1$ and setting $\delta=1$, we obtain
\begin{equation}
r_h=\frac{m}{\alpha}\left(1+\frac{4m^2}{\ell^2{\alpha^2}}+\cdot
\cdot\cdot\right)\left(1+\sqrt{1-\frac{q^2\alpha}{m^2}}\right)\label{eq6},
\end{equation}
which can be written as
\begin{equation}
r_h=\frac{1}{\alpha}\left(1+\frac{4m^2}{\ell^2{\alpha^2}}+\cdot
\cdot\cdot\right)\left(m+\sqrt{m^2-q^2\alpha}\right)\label{eq7},
\end{equation}
where $\alpha=\sqrt{1+\frac{4q^2}{\ell^2}}$.

that is, the event horizon of the RNdS black hole is greater than the Reissner-Nordstr\"{o}m event horizon $r_{RN}=m+\sqrt{m^2-q^2}$

Again it gives the RN black hole \cite{fifty four} horizons for $\ell\rightarrow\infty$ and Schwarzschild-de Sitter black hole horizon for $q=0$. The metric the (\ref{eq1}) represents an interesting asymptotically de-Sitter extreme RN black hole for $q^2=\alpha m^2$, while for $q^2>\alpha m^2$ it does represent any black hole but an unphysical naked singularity
at $r=0$.
We now define $\Delta=r^2+q^2-2mr-\frac{r^4}{\ell^2}$ and then the line element becomes
\begin{equation}
ds^2=-\frac{\Delta}{r^2}dt^2+\frac{r^2}{\Delta}dr^2+r^2(d\theta^2+{\rm sin^2\theta}d\phi^2)\label{eq8}.
\end{equation}
Its position of black hole horizon is same as given in Eq. (\ref{eq7}).

\section{The Hamilton-Jacobi Method}\label{sec3}
In the Hamilton-Jacobi method we avoid the exploration of the equation of motion in the Painlev\'e coordinates system. To calculate the imaginary part of the action from the relativistic Hamilton-Jacobi equation, the action $I$ of the outgoing particle from the black hole horizon satisfies the relativistic Hamilton-Jacobi equation
\begin{equation}
g^{\mu\nu}\left(\frac{\partial I}{\partial x^\mu}\right)\left(\frac{\partial I}{\partial x^\nu}\right)+u^2=0,\label{eq9}
\end{equation}
in which $u$ and $g^{\mu\nu}$ are the mass of the particle and the inverse metric tensors derived from the line element of the near black hole horizon.

The line element near the black hole horizon becomes
\begin{equation}
ds^2=-\frac{\Delta_{,r}(r_h)(r-r_h)}{r^2_h}dt^2+\frac{r^2_h}{\Delta_{,r}(r_h)(r-r_h)}dr^2+r^2_h(d\theta^2+{\rm sin^2\theta}d\phi^2)\label{eq10},
\end{equation}
where,
\begin{equation}
\Delta_{,r}(r_h)=\frac{d\Delta}{dr}\bigg |_{r=r_h}=2(r_h-m-2\frac{r^3_h}{\ell^2}).\label{eq11}
\end{equation}
In the tunneling interpretation of Hawking radiation, utilizing WKB approximation \cite{sixty eight}, the probability of radiation is related to the imaginary part of the tunneling particle as
\begin{eqnarray}
\Gamma \sim {\rm exp}(-2{\rm Im}I)\label{eq12}
\end{eqnarray}
The non-null inverse metric tensors for the metric (\ref{eq10}) are
\begin{eqnarray}
g^{00}=-\frac{r^2_h}{\Delta_{,r}(r_h)(r-r_h)}, \quad g^{11}=\frac{\Delta_{,r}(r_h)(r-r_h)}{r^2_h}, \quad g^{22}=\frac{1}{r_h^2}, \quad g^{33}=\frac{1}{r_h^2{\rm sin^2\theta}}.\label{eq13}
\end{eqnarray}
We can write Eq. (\ref{eq9}) with the help of  Eq. (\ref{eq13}) as
\begin{equation}
-\frac{r^2_h}{\Delta_{,r}(r_h)(r-r_h)}\left(\frac{\partial I}{\partial t}\right)^2+\frac{\Delta_{,r}(r_h)(r-r_h)}{r^2_h}\left(\frac{\partial I}{\partial r}\right)^2+\frac{1}{r_h^2}\left(\frac{\partial I}{\partial \theta}\right)^2+\frac{1}{r_h^2{\rm sin^2\theta}}\left(\frac{\partial I}{\partial \phi}\right)^2+u^2=0\label{eq14}.
\end{equation}
To solve the action $I$ for $I(t, r, \theta, \phi)$. Considering the properties of black hole spacetime, the separation of variables can be taken as follows
\begin{equation}
I=-\omega t+R(r)+H(\theta)+j\phi\label{eq15},
\end{equation}
where $\omega$ and $j$ are respectively the energy and angular momentum of the particle. Since RNdS black hole is nonrotating, the angular velocity of the particle at the horizon is zero. Inserting  Eq. (\ref{eq15}) into Eq. (\ref{eq14}) and solving $R(r)$ holds an expression of
\begin{eqnarray}
R(r)=\pm\frac{r^2_h}{\Delta_{,r}(r_h)}\int \frac{dr}{(r-r_h)}\quad \times \sqrt{\omega^2-\frac{\Delta_{,r}(r_h)(r-r_h)}{r^2_h}\left[g^{22}(\partial_\theta H(\theta))^2+g^{33}j^2+u^2\right]}\label{eq16}.
\end{eqnarray}
For the convenience of research, let's the emitted particle as an ellipsoid shell of energy $\omega$ to tunnel across the event horizon and should not have  motion in  $\theta$-direction ($d\theta=0$). The quadratic form of Eq.(\ref{eq14}) is the reason of $\pm$ signatures that summarized in the above equation. Solution of Eq. (\ref{eq16}) with ``+'' signature corresponds to outgoing particles and the other solution i.e., the solution with ``-''signature refers to the ingoing particles. The solution given by Eq.(\ref{eq16}) is singular at $r=r_h$ which corresponds to the event horizon. Finishing the above integral by using the Cauchy\rq s integral formula, we obtain
\begin{eqnarray}
R(r)=\pm \frac{2\pi i r^2_h}{\Delta_{,r}(r_h)}\omega \label{eq17},
\end{eqnarray}
Substituting the above result in Eq. (\ref{eq15}), the imaginary part of the action I corresponding to the outgoing particle is obtained by $\pi $ times the
residue of the integrand
\begin{eqnarray}
{\rm Im}I =\frac{\pi r^2_h }{r_h-m-2\frac{r^3_h}{\ell^2}}\omega\label{eq18}.
\end{eqnarray}
Using Eq. (\ref{eq7}) into Eq. (\ref{eq18}), we get the imaginary part of action as
\begin{eqnarray}
&&{\rm Im}I
=\frac{\frac{1}{\alpha^2}\left(1+\frac{4m^2}{\ell^2{\alpha^2}}+\cdot
\cdot\right)^2{(m+\sqrt{m^2-q^2\alpha})^2}}{\frac{1}{\alpha}\left(1+\frac{4m^2}{\ell^2{\alpha^2}}+\cdot
\cdot\right){(m+\sqrt{m^2-q^2\alpha})}-m-\frac{2}{\ell^2\alpha^3}\left(1+\frac{4m^2}{\ell^2{\alpha^2}}+\cdot
\cdot\right)^3(m+\sqrt{m^2-q^2\alpha})^3}\omega\label{eq19}\\
&&=\frac{\frac{1}{\alpha^2}{(m+\sqrt{m^2-q^2\alpha})^2}}{\frac{1}{\alpha}\left[\left(1-\frac{4m^2}{\ell^2{\alpha^2}}+\cdot
\cdot\right){(m+\sqrt{m^2-q^2\alpha})}-m\alpha\left(1-\frac{8m^2}{\ell^2\alpha^2}+\cdot\cdot\right)-\frac{2}{\ell^2\alpha^2}\left(1+\frac{4m^2}{\ell^2{\alpha^2}}
+\cdot\cdot\right)(m+\sqrt{m^2-q^2\alpha})^3\right]}\omega\nonumber
\end{eqnarray}
Now for the simplicity, neglecting $m^3$ and its higher order terms, we then get
\begin{eqnarray}
{\rm Im}I  =\frac{1}{\alpha}.\frac{(m+\sqrt{m^2-q^2\alpha})^2}{(m+\sqrt{m^2-q^2\alpha})-m\alpha}\omega\label{eq20}.
\end{eqnarray}

 We now focus on a semiclassical treatment of the associated radiation and adopt the picture of a pair of virtual particles spontaneously created just inside the horizon. The positive energy virtual particle can tunnel out while the negative one is absorbed by the black hole resulting in a decrease in the mass. In presence of cosmological constant, we fix the ADM(Arnowitt-Deser-Misner) mass of the total spacetime and allow the RNdS black hole to fluctuate. When a particle with energy $\omega$ tunnels out, the mass of the RNdS black hole changed into $m-\omega$. Since the angular velocity of the particle at the horizon is zero $(\Omega_h=0)$, the angular momentum is equal to zero. Assuming the self-gravitational interaction into account, the imaginary part of the true action can be calculated from Eq. (\ref{eq18}) in the following integral form
\begin{eqnarray}
{\rm Im}I=\pi \frac{1}{\alpha}.\int^\omega_0\frac{(m+\sqrt{m^2-q^2\alpha})^2}{(m+\sqrt{m^2-q^2\alpha})-m\alpha}d\omega'\label{eq21}
\end{eqnarray}

\begin{eqnarray}
{\rm Im}I=\pi \frac{1}{\alpha}.\int^\omega_0\frac{(m+\sqrt{m^2-q^2\alpha})^2}{\sqrt{m^2-q^2\alpha}+(1-\alpha)m}d\omega'\label{eq22}
\end{eqnarray}
For the maximum value of integration, neglecting $(1-\alpha)m$. Equation (\ref{eq22}) becomes
\begin{eqnarray}
{\rm Im}I=\pi \frac{1}{\alpha}.\int^\omega_0\frac{\left(m+\sqrt{m^2-q^2\alpha}\right)^2}{\sqrt{m^2-q^2\alpha}}d\omega'\label{eq23}
\end{eqnarray}
Replacing $m$ by $m-\omega$ we have
\begin{eqnarray}
{\rm Im}I=-\pi\frac{1}{\alpha}.\int^{(m-\omega)}_m\frac{\left(m-\omega'+\sqrt{(m-\omega')^2-q^2\alpha}\right)^2}{\sqrt{(m-\omega')^2-q^2\alpha}}d(m-\omega')\label{eq24}
\end{eqnarray}
\begin{eqnarray}
{\rm Im}I=-\pi \frac{1}{\alpha}.\int^{(m-\omega)}_m\frac{2(m-\omega')^2+2(m-\omega')\sqrt{(m-\omega')^2-q^2\alpha}-q^2\alpha}{\sqrt{(m-\omega')^2-q^2\alpha}}d(m-\omega')\label{eq25}
\end{eqnarray}
Finishing the integral we get
\begin{eqnarray}
{\rm Im}I=-\pi \frac{1}{\alpha}.[{(m-\omega)\sqrt{(m-\omega)^2-q^2\alpha}+(m-\omega)^2-m\sqrt{m^2-q^2\alpha}-m^2}] \label{eq26}
\end{eqnarray}

Using Eq.(12) the tunneling rate for RNdS black hole is given by
\begin{eqnarray}
\Gamma \sim {\rm exp}(-2{\rm Im}I)&=&{\rm exp}\left\{\pi . \frac{1}{\alpha}\left[2(m-\omega)^2+2(m-\omega)\sqrt{(m-\omega)^2-q^2\alpha}-2m\sqrt{m^2-q^2\alpha}-2m^2\right]\right\}\nonumber\\
&=&{\rm exp}[\pi(r^2_f-r^2_i)]\nonumber\\
&=&{\rm exp}(\Delta S_{BH}),\label{eq27}
\end{eqnarray}
where $r_f=\frac{1}{\sqrt\alpha}[(m-\omega)+\sqrt{(m-\omega)^2-q^2\alpha}]$  and  $r_i=\frac{1}{\sqrt\alpha}[m+\sqrt{m^2-q^2\alpha}]$
are the locations of the RNdS event horizon before and after the particle emission respectively, and $\Delta S_{BH}=S_{BH}(m-\omega)-S_{BH}(m)=\pi(r^2_f-r^2_i)$ is the change of Bekenstein-Hawking entropy.

\section{Purely thermal radiation}\label{sec4}
The radiation spectrum is not pure thermal although gives a correction to the Hawking radiation of RNdS black hole as point out by Eq. (\ref{eq27}). In the form of a thermal spectrum, using the WKB approximation the tunneling rate is also related to the energy and the Hawking temperature of the radiative particle as
$\Gamma \sim {\rm exp}(-\frac{\Delta \omega}{T})$. If $\Delta \omega < 0$ is the energy of the emitted particle then due to energy conservation, the energy of the outgoing shell must be $-\Delta \omega$, then above expression becomes
\begin{eqnarray*}
\Gamma \sim {\rm exp}(\frac{\Delta \omega}{T})
\end{eqnarray*}
Now using the first law of thermodynamics, we can write  $\Gamma \sim {\rm exp}({\Delta S})$, which is related to the change of Bekenstein-Hawking entropy as follows
\begin{eqnarray}
\Gamma \sim {\rm exp}({\Delta S_{BH}})={\rm exp}\{S_{BH}(M-\omega)-S_{BH}(M)\}\label{eq28}
\end{eqnarray}
We establish Eq.(\ref{eq27}) as developed by Rahman et al. \cite{sixty one} in power of $\omega$ upto second order using Taylor\rq s theorem of the form
\begin{eqnarray}
\Gamma \sim {\rm exp}(\Delta S_{BH})={\rm exp}\left\{-\omega \frac{\partial S_{BH}(m)}{\partial m}+\frac{\omega^2}{2}\frac{\partial^2 S_{BH}(m)}{\partial m^2}\right\}
        ={\rm exp}[\pi (-\omega\beta+\frac{\omega^2}{2}\gamma)]\label{eq29}
\end{eqnarray}
where $\beta=\frac{2}{\alpha}\left[2m+\sqrt{m^2-q^2\alpha}+\frac{m^2}{(m^2-q^2\alpha)^{\frac{1}{2}}}
\right]$ and  $\gamma=\frac{2}{\alpha}\left[2+\frac{3m}{\sqrt{m^2-q^2\alpha}}-\frac{m^3}{(m^2-q^2\alpha)^{\frac{3}{2}}}
\right]$ \\
When $\ell \rightarrow \infty$, then $\alpha=1$ the pure thermal spectrum can be reduced for Reissner-Nordstr\"{o}m  black hole \cite{fifty four}. The radiation spectrum given by  (\ref{eq27}) is more accurate and provides an interesting correction to Hawking pure thermal spectrum.

\section{Concluding Remarks}\label{sec5}
 Hawking radiation as massive particle tunneling method from RNdS black hole have been presented in this paper. By taking into account the self-gravitational interaction, the background spacetime as dynamical and the energy as conservation, we have recovered that the tunneling rate at the event horizon of RNdS black hole is related to the Bekenstein-Hawking entropy. Specially, when $\ell \rightarrow \infty$, then $\alpha=1$ the RNdS black hole reduced to Reissner-Nordstr\"{o}m  black hole. The positions of the event horizon of Reissner-Nordstr\"{o}m black hole before and after the emission of the particles with energy $\omega$ are $r_i=m+\sqrt{m^2-q^2}$ and $r_f=(m-\omega)+\sqrt{(m-\omega)^2-q^2}$ and hence the tunneling rate of Reissner-Nordstr\"{o}m black hole can be written as
\begin{eqnarray}
\Gamma \sim {\rm exp}(-2{\rm Im}I)&=&{\rm exp}\left\{\pi\left[\left\{(m-\omega)+\sqrt{(m-\omega)^2-q^2}\right\}^2-\left\{m+\sqrt{m^2-q^2}\right\}^2\right]\right\}\nonumber\\
&=&{\rm exp}[\pi(r^2_f-r^2_i)]={\rm exp}(\Delta S_{BH}).\label{eq30}
\end{eqnarray}
Obviously, when $q=0$ Eq. (\ref{eq27}) gives the result of SdS black hole \cite{sixty one} and when $\ell \rightarrow \infty \,\mbox{and} \,q=0$ our result coincides with that obtained by Parikh and Wilczek \cite{fourty six}.\\

In addition, our discussions made here can be directly extended to the anti-de Sitter case  by changing the sign of the cosmological constant to a negative one, and also can be easily generalized to higher dimensional spherically symmetric black holes case.

{\bf Acknowledgement}\\
Both the authors thanks the Abdus Salam International Centre for Theoretical Physics (ICTP), Trieste, Italy, for giving opportunity to utilize its e-journals for research purpose.

\end{document}